\documentclass[preprint,showpacs,preprintnumbers,amsmath,amssymb]{revtex4}

\usepackage{graphicx}
\usepackage{dcolumn}
\usepackage{bm}

\begin{document}

\preprint{}

\title{
Hadron decay amplitudes from $B\to K \pi$ and $B\to \pi \pi$ decays }

\author{
$^{1,2}$Xiao-Gang He\footnote{hexg@phys.ntu.edu.tw. } and
$^3$Bruce H. J.
McKellar\footnote{mckellar@physics.unimelb.edu.au}}
\affiliation{%
$^1$Department of Physics, Peking University, Beijing\\
$^2$NCTS/TPE, Department of Physics, National Taiwan University,
Taipei\\
$^3$ School of Physics, University of Melbourne, Parkville, Vic
3052}

\date{10 June, 2005}

\begin{abstract}
One can analyze the hadronic decay amplitudes for $B\to PP$ decays
using flavor SU(3) symmetry in different ways, such as the
algebraic and quark diagram approaches. We derive specific
relations between these two sets of amplitudes. In the Standard
Model the leading hadronic decay amplitudes depend on only  five
independent parameters which can be determined using recent
experimental data on the branching ratios and CP violating
asymmetries of $B\to K \pi$ and $B \to \pi\pi$. We find, however,
that the leading amplitudes provide a best fit solution with a
large $\chi^2$, which cannot therefore be regarded as a good fit.
Keeping sub-leading terms, makes it  possible to have a reasonable
minimal $\chi^2$. We also find that in general the color
suppressed decay amplitude is comparable with the color allowed
amplitude, contrary to expectations.
\end{abstract}

\maketitle

\section { Introduction}

Recently the Babar and Belle collaborations have measured direct
CP violation in $\bar B^0 \to K^- \pi^+$ with consistent results
which average to $A_{CP}(K^-\pi^+) = -0.109\pm 0.019$.
\cite{kpicp,ICHEP}. They have also given new,  precision
determinations of the branching ratios of $B\to K \pi$ and $B \to
\pi\pi$, which are compiled in \cite{hfag}, and are given in Table
\ref{datatable}. In the table we also list other CP violating
variables although they are not as precisely measured as the
branching ratios. These results, and those for other two body $B$
decays, show that the study of two body charmless $B$ decays has
entered a precision era. They can be used to understand the
dynamics of $B$ decays and CP violation in the Standard Model.
\begin{table}[htb]
\centering
\begin{tabular}{|c|c|c|c|}\hline
Decay channel & $BR \times 10^6$& $A_{CP}$&$S_f$ \\
\hline $ \bar K^0 \pi^-$ & $24.1\pm
1.3$& $-0.02\pm 0.034$&--\\
\hline $K^- \pi^0$ &
 $12.1\pm 0.8$&$0.04\pm 0.04$&--\\
 \hline
$K^-\pi^+$ &  $18.2\pm 0.8$&$-0.109\pm 0.019$&--\\
\hline
$\bar K^0 \pi^0$ &  $11.5\pm 1.0$&$0.09\pm 0.14$&$0.34\pm 0.28$\\
\hline
$\pi^-\pi^0$ & $5.5\pm 0.6$&$-0.02\pm 0.07$&--\\
\hline
$\pi^-\pi^+$ & $4.5\pm 0.4$&$0.37\pm0.10$&$-0.50\pm 0.12$\\
\hline
$\pi^0\pi^0$ & $1.45\pm 0.28$&$0.28\pm 0.39$&-- \\
\hline
\end{tabular}
\caption{Experimental data on $B\to K \pi, \pi\pi$. The
normalizations of $S_f$  are
 $\sin(2\beta)$ and $\sin(2\alpha)$ in the cases where tree
and penguin amplitudes are neglected for $\bar K^0 \pi^0$ and
$\pi^+\pi^-$, respectively.} \label{datatable}
\end{table}

The decay amplitudes for $B\to K\pi, \pi\pi$ can be parameterized
according to SU(3) (or isospin) symmetry through the equivalent
quark diagram or algebraic representations. In these ways,
detailed in \cite{diagram1,diagram2, algebra1,algebra2}, the decay
amplitudes in these and other two body charmless $B$ decays are
related. With enough information, the parameters can be completely
fixed. These decay amplitudes, or the equivalent parameters, can
also be estimated using various different theoretical approaches.
In this work we will carry out our analysis as model independently
as possible by using flavor symmetries to study the implications
of the measured branching ratios and CP asymmetries in $B\to K
\pi, \pi\pi$ for the hadronic parameters. Were the parameters to
be well determined by the data, one could regard them as features
to be explained by attempts to calculate the hadronic matrix
elements.

There are many recent studies for $B\to K \pi$ and $B\to \pi\pi$
decays emphasizing on the determination of the CKM matrix elements
and implications for new physics beyond the SM using different
approaches\cite{algebra2,buras,wuzhou}. We take a different
approach in this analysis by taking CKM parameters as the known
ones determined from other data\cite{PDG}, and emphases on the
determination of the leading and sub-leading contributions to the
hadronic parameters in the SM using the most recent data. We first
show that different approaches based on SU(3) flavor symmetry are
completely equivalent when appropriate terms are taken into
account, and obtain specific relations for amplitudes used in
diagram and algebraic approaches. We then use available data to
determine related hadronic amplitudes. We find that the leading
amplitudes provide a best fit solution with a large $\chi^2$, more
than 2 per degree of freedom, which cannot be regarded as good
fit.  Keeping the sub-leading terms, makes it possible to fit the
data. We also find that in general the color suppressed decay
amplitude is comparable with the color allowed amplitude.

Since the first draft of this paper, several articles have been
written on related subjects\cite{buras2, wu2,kim}, and similar
results have been obtained although different authors emphases
different features of the analysis.

\section{ Parametrization of the decay amplitudes}

There are several different ways of parameterizing SU(3) decay
amplitudes for $B\to PP$ decays. We start our analysis by showing
that they are all equivalent when the appropriate terms are included.

In the SM, the decay amplitudes for $B\to K \pi, \pi\pi$ can be
parameterized by separating the terms according to the relevant
products of CKM matrix elements:
\begin{eqnarray}
A_{B \to K\pi} &=& V_{ub}V_{us}^* T_{K\pi} - V_{tb}V_{ts}^* P_{K
\pi},\nonumber\\
A_{B \to \pi\pi} &=& V_{ub}V_{ud}^* T'_{\pi\pi} - V_{tb}V_{td}^*
P'_{\pi \pi},
\end{eqnarray}
where $V_{ij}$ are the CKM matrix elements which in general
contain CP violating phases. The amplitudes $T_{PP} (T'_{PP})$ and
$P_{PP} (P'_{PP})$ are hadronic matrix elements which in general
contain CP conserving final state interaction phases. The primed
$(T',P')_{PP}$ and the un-primed $(T,P)_{PP}$ amplitudes are equal
in the flavor SU(3) symmetry limit. In the above we have used the
unitarity of the CKM matrix to eliminate terms proportional to
$V_{cb}V_{cs}^*$ and $V_{cb}V_{cd}^*$ in favor of the above two
terms.

In the SM the quark level Hamiltonian\cite{wilson} $H$, expanded in
dimension six operators,
\begin{eqnarray*}
 H &  = & {\frac{G_F}{ \sqrt{2}}}\bigg[V_{ub}V_{uq}^* (c_1O_1+c_2O_2)-
\sum_jV_{jb}V_{jq}^*\sum^{10}_{i=3}c^j_i O_i\bigg],
\end{eqnarray*}
requires the hadronic matrix elements $T_{PP}$ and $P_{PP}$
transform under SU(3) as $\overline{3}$, 6 and $\overline{15}$. In
the above $c_i$ are the Wilson coefficients of the operators $O_i$
which have been calculated to next-to-leading order in QCD
corrections. We will use the values calculated in the NDR
regularization scheme at $\mu = m_b$ given in Ref.\cite{wilson}.

The amplitude $T_{PP}$ is dominated by the operators $O_{1,2}$,
which generate, in quark diagram language, the ``color allowed''
$T$ and ``color suppressed'' $C$ amplitudes, containing
$\overline{3}$, 6 and $\overline{15}$ irreducible amplitudes. The
$P_{PP}$, called the ``penguin''amplitude, is generated at the
loop level and contains strong and electroweak penguin
contributions. The strong penguin induces only a $\bar 3$
amplitude, but the electroweak penguin, dominated by the operators
$O_{9,10}$, induces all $\bar 3$, $6$ and $\overline{15}$
amplitudes. As far as the SU(3) group structure is concerned, the
electroweak penguin operators are proportional to $3O_{1,2}/2
-(1/2)O_{3,4}$. One can easily obtain the electroweak penguin
irreducible amplitude by grouping the part proportional to
$O_{3,4,5,6}$ into a strong penguin like operator and the rest
into a tree like operator.

The decay amplitudes can be parameterized according to the SU(3)
irreducible amplitudes\cite{algebra2}, separating the
``annihilation'' amplitudes, $A^{T,P}_{\bar 3,\overline{15}}$,  in
which the initial quarks are annihilated in the weak interaction
from the amplitudes $C^{T,P}_{\bar 3, 6, \overline{15}}$, in which
one of the initial quarks is preserved.   There is also an $A_6$
amplitude, which has the same coefficients as the $C_6$ amplitude and
can therefore be absorbed into it.  We list the ``tree''
amplitudes for $B\to K\pi, \pi\pi$ in Table \ref{algeb}.

\begin{table}[htb]
\centering
\begin{tabular}{|l|l|}\hline
Decay Mode& $SU(3)$ Invariant Amplitude\\
\hline
$T_{\pi^-\pi^0}$ & ${8\over \sqrt{2}} C^T_{\overline{15}}$ \\
\hline
 $T_{\pi^+\pi^-}$ & $2 A^T_{\overline{3}} +
A^T_{\overline{15}}
+ C^T_{\overline{3}} + C^T_6 + 3 C^T_{\overline{15}}$\\
\hline

$T_{\pi^0\pi^0}$& $ {1\over \sqrt{2}}(2 A^T_{\overline{3}} +
A^T_{\overline{15}} + C^T_{\overline{3}} + C^T_6 -5
C^T_{\overline{15}})$ \\

\hline
 $T_{\pi^-\bar K^0}$ & $ 3A^T_{\overline{15}} +
C^T_{\overline{3}}
- C^T_6 - C^T_{\overline{15}}$\\

\hline
 $T_{\pi^0 K^-}$&${1\over \sqrt{2}} (3 A^T_{\overline{15}}
+
C^T_{\overline{3}} - C^T_6 + 7 C^T_{\overline{15}})$\\
\hline
 $T_{\pi^+ K^-}$&$- A^T_{\overline{15}}
+ C^T_{\overline{3}} + C^T_6 + 3 C^T_{\overline{15}}$\\
\hline

$T_{\pi^0\bar K^0}$&$ -{1\over \sqrt{2}} (- A^T_{\overline{15}} +
C^T_{\overline{3}} + C^T_6 -5 C^T_{\overline{15}})$\\
\hline
\end{tabular}
\caption{The SU(3) invariant amplitude for $B\to \pi\pi, \pi K$
decays. Similar amplitude for the strong and electroweak penguin
amplitudes.}\label{algeb}
\end{table}

In the quark diagram approach, the decay amplitudes for various
decay modes are parameterized by the $T$, $C$, $P$, $A$, $E$ and
$P_A$ amplitudes which parameterize the color allowed, color
suppressed, the flavor triplet strong penguin, the annihilation,
exchange and penguin annihilation amplitudes. The details for each
decay amplitudes in terms of the above diagram amplitudes can be
found in Ref.\cite{diagram1, diagram2}. In the diagram approach
neglecting annihilation contributions, the amplitude for $B^-\to
\pi^-\bar K^0$ vanishes which implies in the algebraic approach
$\delta^T = C^T_{\overline{3}} - C_6 -C^T_{\overline{15}}=0$.
Comparing the decay modes $B^- \to \pi^- \pi^0$, and $\bar B^0 \to
\pi^+  K^-, \pi^0 \bar K^0$, one can identify, $8V_{ub}V_{us}^*
C_{\overline {15}} = (T+C)e^{-i\gamma}$, and $4V_{ud}V_{us}^*C_6 =
(T-C)e^{-i\gamma}$. After restoring the annihilation (exchange)
contributions, we find the following relations between the
algebraic and diagram amplitudes:
\begin{eqnarray}
&&T e^{-i\gamma} = V_{ub}V_{us}^* (- A_{\overline{15}} +2 C_6 + 4
C_{\overline{15}}),\;\;C e^{-i\gamma} = V_{ub}V_{us}^*
(A_{\overline{15}} - 2 C_6 + 4
C_{\overline{15}})\;,\nonumber\\
&&A e^{-i\gamma} = 3 V_{ub}V_{us}^*
A_{\overline{15}}\;,\;\;\;\;\;\;\;\;\;\; Ee^{-i\gamma} = 2
V_{ub}V_{us}^*(A_{\overline{3}} + A_{\overline{15}})\,.
\end{eqnarray}

With the above relations one finds that the tree contributions,
terms proportional to $V_{ub}V_{us}^*$, are equivalent in form for
the algebraic and diagram approaches.

When $\delta^T$ is not equal to zero, there seems to be a conflict in
that there are five and four independent variables for the
algebraic and diagram amplitudes, respectively. This puzzle is
resolved by realizing that the diagram amplitudes listed above
has missed a piece of contribution, the penguin
contribution with u-quark in the loop (and also a c-quark since we
have used the CKM unitarity to eliminate the term proportional to
$V_{cb}V_{cs}^*$ due to c-quark in the loop). Indicating this
contribution by $P_{cu}$, the charming penguin, and comparing with
the algebraic amplitudes, one can identify $P_{cu} e^{-i\gamma} =
V_{ub}V_{us}^* \delta^T$.

For the strong penguin amplitude, one identifies $P =
V_{tb}V_{ts}^*C^P_{\overline{3}}$, and $P_A = V_{tb}V_{ts}^*
A^P_{\overline{3}}$. When electroweak penguin amplitudes are
included, one can define a set of parameters $T^{EW}$, $C^{EW}$,
$P^{EW}$, $A^{EW}$, $E^{EW}$ and $P_A^{EW}$ similar to the
previously defined tree quark diagram amplitudes. Here there is no
need to introduce an additional electroweak penguin amplitude
analogous to the $P_{cu}$ amplitude because $P^{EW}$
already includes such a contribution.

With the above relations between the parameters used in the
algebraic and diagram approaches, we therefore have shown that the
two ways of parameterizing $B\to \pi \pi, \pi K$, the algebraic
and diagram approaches, are fully equivalent.

In the SM the tree amplitudes and the electroweak penguin
amplitudes are dominated by $O_{1,2}$ and $O_{9, 10}$ (the
operators $O_{7,8}$ have much smaller Wilson coefficients and can
be neglected to a good precision) where the SU(3) invariant
amplitudes $C_6$ and $C_{\overline{15}}$ originate. Since these
operators have the same Lorentz structure and $O_{9,10} = {3\over
2} O_{1,2} -{1\over 3}O_{3,4}$, one
finds\cite{algebra2,rosnerneubert}that $C^P_6 = -(3/2) \kappa^-
C^T_6$ and $C^P_{\overline{15}}(A^P_{\overline{15}}) = (3/2)
\kappa^+C^T_{\overline{15}}(A^T_{\overline{15}})$. Here
$\kappa^{\pm} = (c_9\pm c_{10})/(c_1\pm c_2)$. These relations
enable one to reduce the number of independent decay amplitudes,
but they, in particular the relation between $C_6^T$ and $C^P_6$,
have not been fully exploited  in many of the analyses in the
literature. There is no simple relation between
$C_{\overline{3}}^T (A^T_{\overline{3}})$ and $C_{\overline{3}}^P
(A^P_{\overline{3}})$. Using the above mentioned relations, one
finds that
\begin{eqnarray}
T^{EW} + C^{EW} = {3\over 2}R \kappa^+ (T+C)\;,\;\; T^{EW} -
C^{EW} = {3\over 2}R \kappa^- (T-C)\;,\;A^{EW} = {3\over
2}R\kappa^+ A,
\end{eqnarray}
where $R = |V_{tb}V_{ts}^*/V_{ub}V_{us}^*|$.

Because that the amplitude $A^T_{\overline{3}}$ is not simply
related to $A^P_{\overline{3}}$ by similar relations for
$C_{6,\overline{15}}$, $E$ and $E^{EW}$ cannot be simply related.
However, for the special case with $A_{\overline 3}=0$, $E^{EW} =
{3\over 2} \kappa^+ E$, $A$ is also related to $E$ by
$E=3A/2$.

The three electroweak amplitudes $T^{EW}$, $C^{EW}$ and $P^{EW}$
are usually written in terms of two amplitudes $P_{EW}$ and
$P^C_{EW}$, as they are not independent since all originated from the
same electroweak penguin operators.  To the leading order this relationship is
$T^{EW} = P^{C}_{EW}$, $C^{EW} = P_{EW}$, and $P^{EW} =
-P^{C}_{EW}/3$.

In the SU(3) limit, the amplitudes for $B\to \pi\pi$ can be
obtained from the previous amplitudes by an appropriate
re-scaling, for the tree amplitudes by $(V^*_{ud}/V^*_{us})\approx
1/\lambda$, and for the strong and electroweak penguin amplitudes
by $r=(V_{td}^*/V_{ts}^*)$. We summarize the complete set of
amplitudes in Table \ref{diagramtable}.

As the amplitudes $P_{cu}$, $A$, $E$, $E^{EW}$ and $P_A$ are
expected to be smaller than the other amplitudes, one hopes to
obtain reasonable description of the relevant data 
even with their contributions ignored. With this approximation,
the analysis is tremendously simplified with only five independent
hadronic parameters in the three complex amplitudes, $T,\; C,\; P$
(one phase of these complex parameters can be absorbed into
redefinition of the meson fields). In the following we will first
carry out an analysis using this approximation. We find however
that
these leading amplitudes 
cannot provide a good
description of the data since the resulting fit has too large a
minimal $\chi^2$.

\begin{table} \centering
\begin{tabular}{|c|c|c|c|c|c|c|c|c|c|c|}\hline
$A(PP)$ & $T$& $C$&$P$&$P_{cu}$&$\kappa^+(T+C)$&$\kappa^-(T-C)$&A&E&$E^{EW}$&$P_A$\\
\hline
$A(\bar K^0 \pi^-)$&0&0&-1&-$e^{-i\gamma}$&0&0&-$e^{-i\gamma}$-${3R\over 2}\kappa^+$&0&0&0\\

$\sqrt{2}A(K^- \pi^0)$&-$e^{-i\gamma}$&-$e^{-i\gamma}$
&-1&-$e^{-i\gamma}$&-${3R\over 2}$&0&-$e^{-i\gamma}$-${3R\over 2} \kappa^+$&0&0&0\\

$A(K^- \pi^+)$&-$ e^{-i\gamma}$&0&-1&-$e^{-i\gamma}$&-${3R\over 4}$&${3R\over 4}$&0&0&0&0\\
 $\sqrt{2}A(
\bar K^0 \pi^0)$&0&-$ e^{-i\gamma}$&1&$e^{-i\gamma}$&-${3R\over 4}$&-${3R\over 4}$&0&0&0&0\\
\hline
$A(PP)$ & $T'$& $C'$&$P'$&$P'_{cu}$&$\kappa^+(T'+C')$&$\kappa^-(T'-C')$&$A'$&$E'$&$E^{'EW}$&$P'_A$\\
\hline
 $\sqrt{2}A(\pi^- \pi^0)$&-${1\over \lambda} e^{-i\gamma}$
 &-${1\over \lambda} e^{-i\gamma}$
&0&0&-${3R\over 2}r$&0&0&0&0&0\\
$A(\pi^- \pi^+)$&-${1\over \lambda} e^{-i\gamma} $&0&-r&-${1\over
\lambda}e^{-i\gamma}$&-${3R\over 4}r$&${3R\over 4}r$&0&
-${1\over \lambda}e^{-i\gamma}$&-$rR$&-r\\
 $\sqrt{2}A(
\pi^0 \pi^0)$&0&-$ {1\over \lambda} e^{-i\gamma} $&r&${1\over
\lambda} e^{-i\gamma}$&-${3R\over 4}r$&-${3R\over 4}r$
&0&${1\over \lambda} e^{-i\gamma}$&$r R$&$r$\\
\hline
\end{tabular}
\caption{The quark diagram amplitudes for $B\to K \pi, \pi\pi$. $R
= |V_{tb}V_{ts}^*/V_{ub}V_{us}^*|$, $\lambda =V_{us}^*/V_{ud}^*$
and $r=V^*_{td}/V^*_{ts}$. In the $SU(3)$ limit
$(T,C,P,A,E,E^{EW},PA) = (T',C',P',A',E',E^{'EW},P'_A)$. We have
written the notation in the above $P(P')$ and $PA (P'_A)$ for the
combined strong and electroweak penguin amplitudes $P -
P^C_{EW}/3$ and $P_A + P_A^{EW}$.} \label{diagramtable}
\end{table}

\section{The $B\to K\pi$ data and the Hadronic Parameters}

At present there are 5 well established measurements for $B\to K
\pi$ decays: the four branching ratios and the CP asymmetry in $\bar
B^0 \to K^- \pi^+$. Using these data and taking the CKM matrix
elements determined from various experimental data, one can
determine the hadronic parameters. We take the central values for
the CKM parameters $s_{12} = 0.2243$, $s_{23} = 0.0413$, $s_{23} =
0.0037$ and the CP violating phase $\gamma (\delta_{13})=60^\circ$
given by Ref.\cite{PDG}. For the central experimental data we
obtain two solutions for the amplitudes $C$ and $T$
\begin{eqnarray}
\hspace*{-1cm}1)  && T = 1.018\, e^{3.092i};\;\;
 \quad C = 1.158\, e^{0.0916i}. \nonumber \\
\hspace*{-1cm} 2)
  && T = 1.016\, e^{-2.978 i}; \quad C = 1.154\, e^{0.0070 i}.
\label{solution}
\end{eqnarray}
It is remarkable that there are any solutions. The magnitude of the
amplitudes is almost the same, for the two solutions, while the
phases have changed.

In the analysis we have  normalized the amplitudes  to the
amplitude of $A(\bar K^0 \pi^-)$ and to obtain the physical numbers
they should be multiplied by a
factor $\sqrt{B^{exp}(\bar K^0 \pi) 16\pi m_B \Gamma_{total}^B}$
with $B^{exp} = 24.1\times 10^{-6}$. Since $P$ is determined by
$A(\bar K^0 \pi^-)$, we set $P =1$ in this case. In the
calculations we have also taken into account the $\bar B^0$ and
$B^-$ lifetime, and (later) the $B\to K \pi$ and $B\to \pi\pi$
phase space differences.

Note that, in each of the solutions, $C$ and $T$ are almost real.
This fits with the intuition that the final state mesons have
large energies and the final state interactions which generate the
phase in $C$ and $T$ would be expected to be weak, leading to
small phases. However the ratios $|C/T|$, $|T/P|$ and $|C/P|$ are
of order one, which was also found in refs~\cite{buras,wuzhou},
and which is in contradiction with expectations from various
theoretical calculations. This poses a problem for our ability to
provide a theoretical basis for the observed amplitudes.

Using the hadronic parameters determined above, one can predict
the CP asymmetries in other $B\to K\pi$ decays. Since $B^-\to \bar
K^0 \pi^-$ has only a $P$ amplitude, no CP asymmetry can be
generated in this decay. There are non-zero asymmetries in the
other two decays. We find that for solution 1), $A_{CP}(K^- \pi^0)
= 0.267$, $A_{CP}(\bar K^0 \pi^0) = -0.006$, $S_{\bar K^0 \pi^0}
=-0.375$. For solution 2), $A_{CP}(K^- \pi^0) = -0.266$,
$A_{CP}(\bar K^0 \pi^0) = -0.198$, $S_{\bar K^0 \pi^0} = -0.378$.
These values are different to the central values of the data and
the two solutions can be distinguished in the near future. We note
that the predicted CP asymmetry for $\bar B^0 \to K^- \pi^0$ for
both solutions are much larger in size than the central value of
the data. This also poses another potential problem for the
solutions.

One can also include the direct CP asymmetry data into a $\chi^2$
fit. If we use all $B\to K \pi$ data, the best fit values for the
parameters and the resulting branching ratios (in unit $10^{-6}$)
and CP asymmetries are given by
\begin{eqnarray}
&&P = 0.999,
\quad T = 0.127\, e^{-0.533 i},\quad C = 0.260\, e^{0.295 i}.\nonumber\\
 &&B(\bar K^0 \pi^-)= 24.06, \quad  B(K^- \pi^0)= 12.30,\nonumber \\
 &&B(K^- \pi^+)
= 10.41, \quad
B(\bar K^0 \pi^0)= 18.47;\nonumber\\
&&A_{CP}(K^- \pi^+) = -0.104,\;\;\;\;\quad  \hspace*{-16pt}
A_{CP}(K^-
\pi^0) = 0.019,\nonumber \\
&&A_{CP}(\bar K^0 \pi^0) = -0.150, \quad  S_{\bar K^0 \pi^0} =
0.339. \label{solution1}
\end{eqnarray}
Note that here $P$ is also a fitting parameter. The $\chi^2_{min}$
is 4.68 for 4 degrees of freedom which represents a reasonable
fit.

If all the CP asymmetries, direct and time-dependent, are measured
to a good precision in the future, one can also check the
consistency of CKM parameters determined from other data by taking
$\rho$ and $\eta$ as unknown and letting them be determined
from $B\to K\pi$ data\cite{algebra2,london}.

In fact the analysis carried out above using the well measured
branching ratios and CP asymmetry  in $K^- \pi^+$ mode already
constrains the allowed range for $\gamma$ since for certain values
of $\gamma$, there are no solutions for the hadronic parameters.
For example $\gamma$ in the interval between $54.5^\circ$ to
$100^\circ$ is allowed, but the intervals between $38.5^\circ$ to
$54.5^\circ$ and $105^\circ$ to $145.5^\circ$ are not allowed.

The above analysis shows that the leading SU(3) amplitudes can
provide a good description of $B\to K \pi$ data if one allows a
larger than expected $|C/T|$ ratio.

\section{ Predictions and Problems for $B\to \pi\pi$ decays}

In the SU(3) limit, once the parameters $T$, $C$ and $P$ have been
determined from $B\to K \pi$ decays, predictions can be made for
$B \to \pi\pi$ decays. An interesting prediction is the CP
asymmetry in $\bar B^0 \to \pi^- \pi^+$ which can be made without
knowing the specific values of the amplitudes $T$, $C$ and $P$.
From Table \ref{diagramtable} and the fact that in the SM
$Im(V_{ub}V_{us}^*V_{tb}^*V_{ts}) =
-Im(V_{ub}V_{ud}^*V_{tb}^*V_{td})$, one has: $ \Delta^{\bar
B^0}_{K^-\pi^+} = -\Delta^{\bar B^0}_{\pi^-\pi^+}$. Here
$\Delta^B_{PP} = \Gamma^B_{PP} - \Gamma^{\bar B}_{\bar P\bar P}$.
This relation leads to\cite{deshhe}
\begin{eqnarray}
A_{CP}(\pi^-\pi^+) = (-1)A_{CP}(K^-\pi^+){B( K^-\pi^+)\over
B(\pi^-\pi^+)}. \label{predict}
\end{eqnarray}
There are $SU(3)$ breaking effects which modify the above
relation. Using the QCD factorization method to make  an estimate
about the SU(3) breaking effects due to meson decay constants and
light cone distribution for different mesons, the factor $-1$ in
the above equation is changed to\cite{deshhe} $-0.9$.  One would
predict a CP asymmetry in $\bar B^0 \to \pi^-\pi^+$ of $0.39\pm
0.08$ which is consistent with the experimental value\cite{hfag}.

We look forward to a high precision measurement of
$A_{CP}(\pi^-\pi^+)$, which will provide a very direct test of the
SM.

Given the values of $T$, $C$ and $P$ more predictions can be made
for $B\to \pi\pi$ decays. However, the values obtained in eqs.
(\ref{solution}) and (\ref{solution1}) would imply branching
ratios of $B^- \to \pi^-\pi^0$, $\bar B^0\to \pi^-\pi^+,
\pi^0\pi^0$ which are too large compared with experimental data.
This is mainly due to the $T$ and $C$ parameters, as determined
from  the $B\to K\pi$ data, being too large for the $\pi\pi$
decays. The simple parametrization and the present experimental
data are not consistent using the same set of leading amplitudes
to explain both the $B\to K \pi$ and $B\to \pi\pi$ data.

The use of central values to determine the hadronic parameters is, of course,
too restrictive. It is possible that in the  ranges allowed by the
experimental errors, a consistent solution can be found. We
therefore carried out a $\chi^2$ fit to determine the parameters.
We consider the case with SU(3) symmetry taking the four $B\to K
\pi$, the three $\pi \pi$ branching ratios, and the direct CP
asymmetry $A_{CP}(\bar B^0 \to K^- \pi^+)$ as the input data
points to determine the best fit values for the five hadronic
parameters. We obtain
\begin{eqnarray}
P = 0.971, \quad T = 0.090\, e^{-2.44 i}, \quad C = 0.076\,
e^{-1.54 i}. \label{ssss}
\end{eqnarray}
We find that with this set of parameters the resulting branching
ratios and CP asymmetry are within the two standard deviation
ranges of the data. However, the $\chi^2$ at the minimum is 8.42
for 3 degrees of freedom which is rather high. This indicates that
there is a potential problem for the leading parametrization to
explain all $B\to K \pi,\;\pi\pi$ data

Including all the  branching ratio and  CP asymmetry data in Table
I in the fitting, we obtain
\begin{eqnarray}
P = 0.941,\quad T = 0.086\, e^{-2.585 i}, \quad C = 0.0817\,
e^{2.65 i}. \label{ssssss}
\end{eqnarray}
With this set of hadronic parameters, the branching ratios and
asymmetries are
\begin{eqnarray}
&&B(\bar K^0 \pi^-)=21.32, \;\;B(\pi^-\pi^0)=4,34, \nonumber\\
&&B(K^- \pi^0)=11.19,\;\;A_{CP}(K^-\pi^0) = -0.015,\nonumber\\
&&B(\bar K^0\pi^0)= 9.753,\;\;A_{CP}(\bar
K^0\pi^0) = -0.053\nonumber\\
&&B(K^-\pi^+)=20.13,\;\;
A_{CP}(K^-\pi^+) = -0.097,\nonumber\\
&&B(\pi^0\pi^0)=1.61,\;\; A_{CP}(\pi^0\pi^0)
= 0.410,\nonumber\\
&&B( \pi^-\pi^+)=4.88,\;\; A_{CP}(\pi^-\pi^+) = 0.388,\nonumber\\
&&S_{\bar K^0 \pi^0} = 0.691,\;S_{\pi^+\pi^-} = -0.78.
\label{sssssss}
\end{eqnarray}
The $\chi^2_{min}$ increased to about 26.5 for 9 degrees of
freedom. This is similar to the fitting quality for the case of
eq. (ssss).

We conclude that the leading order parametrization has a
 problem
with the present data. That could be due to the smaller
sub-leading terms of the SM which we neglected in constructing the
leading amplitudes parametrization playing a more important role
than expected, or to $SU(3)$ breaking effects. It may be due to
the quality of the data. And it may also be due to physics beyond
the SM\cite{buras,wuzhou,newp}. Before making any claims for the
existence of new physics beyond the SM, one must make sure that
the SM contributions, leading and also sub-leading included,
cannot account for the data. In the next section we analyze the
effects of the sub-leading terms which have been neglected in the
previous analysis.

\section{Expanding the Parameter Set}

A number of approximations have been made to obtain the leading
amplitude parametrization with just five independent hadronic
variables: 1) $\delta^T = 0$, 2) neglect of the annihilation
amplitude $Ae^{-i\gamma}= 3V_{ub}V_{us}^*A^T_{\overline{15}}$, the
exchange amplitude $Ee^{-i\gamma}
=2V_{ub}V_{us}^*(A^T_{\overline{3}} + A^T_{\overline{15}}) $, and
penguin annihilation amplitude $P_A = A^P_{\overline {3}}$, and 3)
SU(3) symmetry hold for the hadronic matrix elements in $B\to PP$
decays.

The first and the second approximations can be tested
experimentally to some degree. If $\delta^T=0$ and $A=0$, CP
asymmetry in $B^- \to \bar K^0 \pi^-$ will be very small. Since
the tree amplitude for $B^-\to K^- K^0$ has the same form as that
for $B^-\to \bar K^0 \pi^-$, the assumption of $\delta^T=0$ and
$A=0$ implies a very small branching ratio for $B^-\to K^-K^0$
(penguin contribution to this decay mode is suppressed).
Experimentally, a non-zero CP asymmetry for $B^- \to \bar K^0
\pi^-$ and a non-zero branching ratio for $B^-\to K^-K^0$ decay
has not been established, $\delta^T \ne 0$ and $A \ne 0$ are not
required. The smallness of $E$ can be tested by $\bar B^0\to K^+
K^-$ since its tree amplitude is proportional to $E$. At present a
non-zero amplitude for this mode has not been established either.
We however find that the experimental upper bounds obtained for
these decays can constrain the parameters $A$, $E$ and $P_{cu}$.
The tree amplitude for $\bar B^0 \to \bar K^0 K^0$ also depends on
$P_{cu}$ and other small parameters, the upper bound on this mode
therefore can also provide constraint on the parameters.  It is
not possible to have a good test of the smallness of $P_A$
directly in $B\to KK$ modes since it is suppressed by a factor of
$\lambda r$. We now analyze whether the restoration of these small
amplitudes can improve the fit.

The contributions from $A$, $E$, $E^{EW}$ and $P_A$  are
annihilation in nature, their sizes are expected to be smaller
than the amplitudes $T$, $C$ and $P$. If the final results from
fit ended up with comparable size for these parameters, one should
not regard the fit a good one. The parameter $P_{cu}$ is a penguin
amplitude in nature, it should be compared with the amplitude
$\sim \lambda^2 P$, where the pre-factor $\lambda^2$ takes care of
the CKM suppression of $P_{cu}$ compared with $P$.

To study the effects of $A$ and $E$, we use two independent
parameters $\epsilon$ and $\tau$ defined as $ \tau e^{-i\gamma}=
V_{ub}V_{us}^* A^T_{\overline{15}}$ and $\epsilon e^{-i\gamma} =
V_{ub}V_{us}^* A^T_{\overline{3}}$ which have definitive SU(3)
irreducible structure.

We find that the minimal $\chi^2$ can be improved. To have
specific idea about how these new sub-leading parameters affects
the decays, we studied three cases with only one of the parameters
$P_{cu}$, $\tau$ and $\epsilon$ to be non-zero separately, and
fitting the measured branching ratios for $B\to K \pi, \pi\pi$ and
direct CP asymmetries for $B\to K^+ \pi^-, \bar K^0 \pi^0, K^-
\pi^0, \pi^+\pi^-, \pi^0\pi^0$. We also include information about
the branching ratios for\cite{hfag} $B\to K^- K^0 (< 2.4\times
10^{-6})$, $\bar K^0 \bar K^0  ((1.19 \pm 0.4)\times 10^{-6})$,
$K^- K^+ (< 0.6\times 10^{-6})$ into the fit. For these modes, the
leading amplitudes are dominated by tree amplitudes with $A(\bar
K^0 K^-) \approx -(3\tau + P_{cu})e^{-i\gamma}/\lambda$, $A(\bar
K^0 K^0) \approx -(2\epsilon - 3 \tau +
P_{cu})e^{-i\gamma}/\lambda$, and $A(K^- K^+\approx -2(\epsilon +
\tau) e^{-i\gamma}/\lambda$. For the two modes having just upper
bounds, we treat their central values to be zero and taking the
68\% c.l. range as errors in the fit. We then predict the values
for $S_{\pi^0\bar K^0}$, $S_{\pi^+\pi^-}$ and $A_{CP}(\bar K^0
\pi^0)$.

Without the sub-leading parameters, the $\chi^2_{min}$ without the
$B\to K K$ data would be about 26. Such a fit cannot be considered
to be a good fit. With the sub-leading parameters, $B\to KK$ can
happen. Including information on $B \to KK$ branching ratios into
the fit, we fid that the $\chi^2_{min}$ are 12, 8, 17 for: i) Only
$P_{cu} \ne 0$; ii) only $\tau \ne 0$; and iii) $\epsilon \ne 0$.
The $\chi^2_{min}$ is significantly reduced. In each of the
fitting above the degrees of freedom is 8. The cases i) and ii)
can be regarded as reasonable fits. We list the best fit values
for the relevant quantities for the cases i) and ii) in the
following.

For i), we have
\begin{eqnarray}
&&P= 0.953,\;\; T = 0.135e^{-2.806i},\;\; C = 0.061e^{1.924i},
\;\;P_{cu} = 0.050 e^{0.057i},\;\;\tau = 0\;\;\;\epsilon = 0\;,\nonumber\\
&&B(\bar K^0 \pi^-)=23.11, \;\;B(\pi^-\pi^0)=5.32, \nonumber\\
&&B(K^- \pi^0)=12.04,\;\;A_{CP}(K^-\pi^0) = 0.027,\nonumber\\
&&B(\bar K^0\pi^0)= 9.31,\;\;A_{CP}(\bar
K^0\pi^0) = -0.083\nonumber\\
&&B(K^-\pi^+)=19.32,\;\;
A_{CP}(K^-\pi^+) = -0.098,\nonumber\\
&&B(\pi^0\pi^0)=1.34,\;\; A_{CP}(\pi^0\pi^0)
= 0.751,\nonumber\\
&&B( \pi^-\pi^+)=4.59,\;\; A_{CP}(\pi^-\pi^+) =0.397 ,\nonumber\\
&&B(\bar K^0 K^0) = 1.26\;,\;\;B(K^- K^0) = 1.26\;,\;\; B(K^- K^+)
\approx 0\;,\nonumber\\
 &&S_{\bar K^0 \pi^0} =0.713 ,\;S_{\pi^+\pi^-} =
-0.874,\;A_{CP}(\bar K^0 \pi^-) = 0.005,
\end{eqnarray}
and for case ii), we have
\begin{eqnarray}
&&P= 0.928,\;\; T = 0.111e^{-0.487i},\;\; C = 0.071e^{0.659i},
\;\;P_{cu} = 0,\;\;\tau = 0.017e^{2.81i}\;\;\;\epsilon = 0\;,\nonumber\\
&&B(\bar K^0 \pi^-)=23.80, \;\;B(\pi^-\pi^0)=5.70, \nonumber\\
&&B(K^- \pi^0)=11.55,\;\;A_{CP}(K^-\pi^0) = 0.017,\nonumber\\
&&B(\bar K^0\pi^0)= 9.86,\;\;A_{CP}(\bar
K^0\pi^0) = -0.087\nonumber\\
&&B(K^-\pi^+)=19.17,\;\;
A_{CP}(K^-\pi^+) = -0.110,\nonumber\\
&&B(\pi^0\pi^0)=1.36,\;\; A_{CP}(\pi^0\pi^0)
= 0.669,\nonumber\\
&&B( \pi^-\pi^+)=4.36,\;\; A_{CP}(\pi^-\pi^+) =0.370 ,\nonumber\\
&&B(\bar K^0 K^0) = 1.27\;,\;\;B(K^- K^0) = 0.06\;,\; B(K^- K^+) =
0.03,\nonumber\\
 &&S_{\bar K^0 \pi^0} =0.626 ,\;S_{\pi^+\pi^-} =
0.504,\;A_{CP}(\bar K^0 \pi^-) = 0.027.
\end{eqnarray}

In the above two cases a small CP asymmetry for the mode $B^- \to
\bar K^0 \pi^-$ is developed because the best fit values for
$P_{cu}$ and $\tau$ are complex. The predicted time dependent CP
asymmetry $S_{\pi^+\pi^-}$ are opposite in sign with case i)
having the same sign as the present experimental data. The two
types of solutions are be easily distinguished by a definitive
measurement of $S_{\pi^+\pi^-}$.

The numerical values obtained for the cases i) and ii) are within
expectations that $\tau$ is smaller than $T$, and $P_{cu}$ is of
order $\lambda^2 P$. We note that the above cases the ratio for
$|C/T|$ is still large which may be completely due to low energy
hadronic physics with in the SM. We conclude that when sub-leading
contributions are included, SM can provide a good fit to $B\to K
\pi$ and $B\to \pi\pi$ data. No new physics beyond the SM is
called for at present. It is obvious that if the small parameters
are simultaneously kept non-zero, better fit can be obtained.

\section{SU(3) breaking effects}

We have carried out all the above analysis with the assumption of
SU(3) symmetry. With SU(3) breaking effects taken into account the
set of parameters for $B\to K \pi$ and $B\to \pi\pi$ can be
different. In the algebraic and diagram approaches described
earlier, the SU(3) breaking effects can be systematically included
by inserting at appropriate places the quark mass terms which will
introduces many new unknown parameters. We will not attempt to
carry out a general analysis here, but to simplify the problem by
assuming that the $B\to K \pi$ amplitudes are scaled by a factor
$f_K/f_\pi$ compared with $B\to \pi\pi$ amplitudes.

We argue that the above re-scaling factor should have taken into
account the leading effects of SU(3) breaking. An intuitive
picture can be obtained from pQCD calculations of the decay
amplitudes. In pQCD calculations decay amplitudes proportional to
$f_B f_\pi f_\pi$ and $f_B f_\pi f_K$ for $B\to \pi\pi$ and $B\to
K \pi$ decays, and therefore the re-scaling factor is $f_K/f_\pi$.
There are other places where SU(3) breaking effects may come from,
for example, the difference between the light-cone distribution
amplitudes for the pion and kaon. Neglecting the later type of
SU(3) breaking effects, the analysis can be carried out in the
same way as previously done since no new adjustable parameter is
introduced.

Using the same data points as what used for the analysis obtaining
eq.(\ref{ssss}), we find that the fit improved slightly with
$\chi^2_{min}$ reduced to 7.3 from 8.4. The best fit values for
the hadronic parameters are
\begin{eqnarray}
P=0.978\;,\;\; T = 0.119 e^{-2.60i}\;,\;\; C = 0.100
e^{-1.57i}\;,\nonumber\\
\end{eqnarray}

Using all available data from $B\to K \pi,\; \pi\pi$,
$\chi^2_{min}$ is reduced to 22.4 from 26.5 with
\begin{eqnarray}
P=0.944\;,\;\; T = 0.116 e^{-2.70i}\;,\;\; C = 0.103e^{2.496i}\;.
\end{eqnarray}

The above analysis shows that the simple re-scaling on the decay
amplitudes  can improve the fit, but cannot completely solve the
problem, and still allows for sub-leading amplitudes to play an
important role.

An extreme SU(3) breaking scenario is the case where $B\to K \pi$
and $B\to \pi\pi$ decays are treated independently. We have seen
that if one just fits $B\to K \pi$ data, the leading amplitudes
can account for data, but predict too large branching ratios for
$B\to \pi\pi$ decays. We now study the consequences of just
fitting the $B\to \pi\pi$ data using the leading amplitudes. We
find that the five parameters $T'$, $C'$ and $P'$ have no problem
in fitting the data on branching ratios. To determine the
parameters, at least two more points of experimental data are
needed which we choose to take the direct and time dependent CP
asymmetries $A_{CP}$ and $S_{\pi\pi}$ for $B\to \pi^-\pi^+$. With
the central values of the data, we find two solutions
\begin{eqnarray}
\hspace*{-0.7cm} 1) \; P = 0.576,& T = 0.098\, e^{-2.17 i},& C =
0.077\, e^{-1.12 i},
;\nonumber\\
\hspace*{-0.7cm} 2) \; P = 0.506,& T  = 0.098\, e^{-2.21 i},& C =
0.083\, e^{2.73 i} .\label{pipi}
\end{eqnarray}

These solutions predict $A_{CP}(\pi^0\pi^0)$ to be $-0.60$ and
$0.18$, respectively, and the second solution is closer to the
experimental central value.

These parameter values are quite different to those in eq.
(\ref{solution}) and in eq. (\ref{solution1}), and if we apply
them to $B\to K\pi$, the branching ratios obtained for $B\to K
\pi$ are too small.

We also carried out another alternative fit by using all $B\to
\pi\pi$. This fit obtains $\chi^2_{min}=0.34$ with
\begin{eqnarray}
P=0.487\;,\;\; T = 0.099 e^{-2,23i}\;,\;\;C = 0.084 e^{2.88i}\;.
\end{eqnarray}

This set of parameters are similar to that obtained in eq.
(\ref{pipi}). But very different than that obtained from using
$B\to K\pi$ data only.

\section{Conclusions}

SU(3) flavor symmetry can simplify the analysis for $B\to K\pi$
and $B\to \pi\pi$. To the leading order there are only five
independent hadronic parameters for these decays in the Standard
Model. The leading amplitudes $T$, $C$ and $P$ can provide a
reasonable description of $B\to K\pi$ decays, but with an
inexplicably large value of the ratio $|C/T|$. When combined with
$B\to \pi\pi$ data, there are more difficulties. One of the
problems is that the ratio $|C/T|$ is still of order one, and much
larger than theoretical estimates. And another is that these
parameters can not give a better than a two standard deviation fit
to the current data.

As a first step in trying to resolve these difficulties, we
studied several possible ways of relaxing approximations made in
the simple parametrization, including sub-leading order terms,
annihilation, exchange and charming penguin amplitudes, and also
SU(3) breaking effects. We find that the inclusion of smaller
sub-leading terms can improve the fit to a reasonable range
although still results a ratio of order one for $|C/T|$ which may
be due to low energy hadronic physics within the Standard Model.
It is too earlier to claim the need of new physics beyond the
Standard Model to explain the $B\to K\pi$ and $B\to \pi\pi$ data.

Finally we would like to make a comment on the source for the
large $\chi^2_{min}$ for the leading parametrization from data
quality point of view. We find that the data point $B(K^- \pi^+)$
makes the largest contribution to $\chi^2$ in the fitting using
both $K \pi$ and $\pi\pi$ data. If one removes this data point,
the $\chi^2_{min}$ for the two cases resulting in eqs.
(\ref{ssss}) and (\ref{ssssss}) would drop down to $0.05$ and
$11$, respectively. This is because that the experimental
branching ratio for $K^- \pi^+$ is smaller than expected from
penguin dominance in $K\pi$ decays. The point we would like to
emphases here is that it is necessary to have more precise data to
help deciding whether the Standard Model can explain both $B\to
K\pi$ and $B\to \pi\pi$.

\section*{Acknowledgements}
XGH would
like to thank H.-Y. Cheng and H.-N. Li for useful discussions.
This work was supported in part by the NSC of ROC and the ARC of
Australia.
\\

\end{document}